\documentclass[twocolumn,showpacs,preprintnumbers,amsmath,amssymb,floatfix]{revtex4}

\usepackage{graphicx}
\usepackage{dcolumn}
\usepackage{bm}

\newcommand{\ApJL}{Astrophys. J. Lett.~}
\newcommand{\ApJS}{Astrophys. J. Supp.~}
\newcommand{\ApJ}{Astrophys. J.~}
\newcommand{\PRL}{Phys. Rev. Lett.~}
\newcommand{\PRD}{Phys. Rev. D~}
\newcommand{\MNRAS}{Mon. Not. R. Astron. Soc.~}

\newcommand{\AnnAsAs}{Ann. Rev. Astron. Astrophys.~}
\def\lcdm  {\rm $\Lambda$CDM\ }

\def\mpc{{\rm\,Mpc}}

\def\vol#1  {{{#1}{\rm,}\ }}

\def\etal{{\it et al.}\ }

\newcount\refno
\newcount\rfno
\def\eq{$^{\the\refno\ }$\advance\refno by 1}
\def\ad{\advance\rfno by 1}

\def\clock{\count0=\time \divide\count0 by 60
     \count1=\count0 \multiply\count1 by -60 \advance\count1 by \time
     \number\count0:\ifnum\count1<10{0\number\count1}\else\number\count1\fi}

\def\myputfigure#1#2#3#4#5%
{\hskip0.03\textwidth\vskip#5pt
\makebox[0pt]{\hskip#2in
\includegraphics[width=#3\textwidth]{#1}}\vskip#4pt\hfill}

\def\Gcm2{\rm G~cm^2}

\def\beq{\begin{equation}}
\def\eeq{\end{equation}}
\def\bea{\begin{eqnarray}}
\def\eea{\end{eqnarray}}

\def \date         {\ifcase\month \message{zero} \or
                    January \or February \or March \or April \or May \or June
                    \or July \or
                    August \or September \or October \or November \or
                    December \fi
                    \space\number\day, \number\year}

\begin{document}


\title{Integrated Sachs-Wolfe effect in Cross-Correlation: \\ The Observer's Manual}

\author{Niayesh Afshordi}
\email{afshordi@astro.princeton.edu}
\affiliation{Princeton University Observatory, Princeton University, Princeton, NJ 08544}%


\date

\begin{abstract}
   The Integrated Sachs-Wolfe (ISW) effect is a direct signature of the
   presence of dark energy in the universe, in the absence of spatial curvature.
   A powerful method for observing the ISW effect is through cross-correlation of the Cosmic
   Microwave Background (CMB) with a tracer of the matter in the
   low redshift universe. In this paper, we describe the
   dependence of the obtained cross-correlation signal on the
   geometry and other properties of a survey of the low redshift universe.
   We show that an all-sky survey with about 10 million
   galaxies, almost uniformly distributed within $0<z<1$ should yield a near optimal ISW
   detection, at $\sim 5\sigma$ level. In order to achieve this level of signal-to-noise,
   the systematic anisotropies in the survey must be below $\sim 0.1 \%$, on the scale of
   $\sim 10$ degrees on the sky, while the systematic error in redshift estimates must
   be less than $0.05$.

   Then, we argue that, while an ISW
   detection will not be a good way of constraining
   the conventional properties of dark energy, it could be a valuable
   means of testing alternative theories of gravity on large physical scales.
\end{abstract}

\maketitle

\section{Introduction}
   One of the fundamental pieces of our understanding of modern
   cosmology comes from the study of anisotropies in the Cosmic Microwave Background (CMB).
   First discovered by the DMR experiment on the COBE satellite in the early 90's
   \citep{smoot}, the observations of CMB anisotropies matured through various
   ground-based/balloon-borne experiments (see \citep{WMAP} for a list) until its
   latest climax by the first year data release of the observations of the WMAP satellite
   \citep{WMAP} in 2003, a cosmic variance limited map of the CMB sky with
   a resolution of $\lesssim 0.5$ degs.

   While most of the fluctuations seen by WMAP and other CMB
   experiments were generated at $z\sim 1000$, low redshift ($z<20$)
   physics generates additional fluctuations \citep{Hu}. These
   secondary anisotropies can be detected through cross-correlating
   CMB measurements with large scale structure observations \citep{turok,ISWs}.
   Following the WMAP data
   release, various groups \citep{crosses} claimed a possible observation of such
   correlation with various galaxy surveys, although at a small
   ($2-3\sigma$) significance level. Most recently, \citep{ALS}
   claimed a correlation between WMAP maps and the 2MASS galaxy
   catalog, at both large and small angles.

   In this paper, we focus on such correlations on angles larger than a few degrees,
   and their only known
   cosmological source, the Integrated Sachs-Wolfe (ISW) effect \citep{SachsWolfe1967}.
   The purpose of this work is to provide comprehensive overview of various observational aspects and
   systematic
   limitations of an ISW detection in cross-correlation, in a flat universe. Secs. II and III will briefly
   review the physics of the expected signal and error in any ISW
   detection. In Sec. IV, we consider over what
   redshift range, and what angular scales, most of the ISW signal arises,
   and what the optimum signal-to-noise for an ISW detection may
   be. Sec. V deals with the limitations of realistic surveys,
   i.e. the Poisson noise, systematic contaminations, and redshift
   errors. Finally, in Sec. VI, we discuss what we
   may (or may not) learn from a detection of the ISW signal in
   cross-correlation, and Sec. VII concludes the paper.

   Throughout this paper, unless mentioned
   otherwise, we use the flat 1st year WMAP+CBI+ACBAR+2dF+Ly-$\alpha$
   concordance cosmological model
   ($\Omega_m = 0.27, \Omega_{\Lambda} = 0.73$, and $\sigma_8 = 0.84$ \citep{SpergelEtAl2003}),
   with a cosmological constant (i.e. $w=-1$), and a running spectral
   index ($n_s=0.93$, and $dn_s/d\ln k = -0.031$ at $k = 0.05 ~h/\mpc$). However, note that our conclusions are not sensitive to
   the details of the model.

    In this paper, the 3D Fourier transform is defined as
    \beq
    A_{\bf k} = \int {\bf d^3 x}~ A({\bf x}) \exp(-i{\bf
    k.x}),
    \eeq
    while the cross-power spectrum of fields $A$ and $B$, $P_{A,B}$,
    is defined through
    \beq
     {\rm Re} \langle A_{\bf k} B^*_{\bf k^{\prime}} \rangle = (2\pi)^3 {\bf
     \delta^3_D(k-k^{\prime})} P_{A,B}(k),
     \eeq
    where ${\bf \delta^3_D}$ is the 3D Dirac delta function. The auto-power spectrum
    is also defined in a similar way.

\section{The ISW effect in Cross-Correlation}
  The ISW effect is caused by the time variation in the cosmic gravitational
  potential, $\Phi$, as CMB photons pass through it. In a flat universe, the
  anisotropy in CMB photon temperature
  due to the ISW effect\citep{SachsWolfe1967} is an integral over the conformal time
  $\eta$
  \bea
  \delta_{\rm ISW}({\bf \hat{n}}) = \frac{\delta T_{\rm ISW}}{T} = 2 \int^{\eta_0}_{\eta_{\rm LS}} \Phi^{'}[(\eta_0-\eta){\bf \hat{n}}, \eta] ~d\eta
  ,\label{isw}
  \eea
  where $\Phi^{'} \equiv \partial \Phi/\partial \eta$, and ${\bf
  \hat{n}}$ is the unit vector along the line of sight. $\eta_{\rm
  LS}$ and $\eta_0$ are the conformal times at the last
  scattering and today, while the speed light is set equal to unity ($c=1$).
  The linear metric is assumed to be
  \beq
  ds^2=a^2(\eta)\{[1+2\Phi({\bf x},\eta)]d\eta^2-[1-2\Phi({\bf x},\eta)]{\bf dx \cdot dx}
  \},\label{metric}
  \eeq
  (the so-called longitudinal gauge) and $\eta_0$ is the conformal time
  at the present.

  In a flat universe, in the linear regime, $\Phi$ does not change with time at any given comoving point for a
  fixed equation of state \footnote{This statement is strictly valid only for wavelengths much larger
  than the Hubble radius, or if the speed of sound vanishes.}, and therefore
  observation of an ISW effect is an indicator of a change in the equation
  of state of the universe. Assuming that this change is due to an
  extra component in the matter content of the universe, the so-called dark energy,
  this component should have a negative pressure to become important at
  late times \citep{ratra}. Therefore, observation of an ISW effect in a flat universe is a
  signature of dark energy.

  The ISW effect
  is observed at large angular scales because most of the power in
 the fluctuations of $\Phi$ is at large scales. Additionally, the
 fluctuations at
  small angles tend to cancel out due to the integration over the
  line of sight.

   We are interested in finding the cross-correlation of the ISW effect with the galaxy distribution.
   Assuming Gaussian initial conditions, and full-sky coverage, different harmonic multipoles are statistically
   independent in the linear regime. Therefore, as the ISW effect is only important on large scales which
   are still linear, the statistical analysis is significantly
   simplified in harmonic space. For a galaxy survey with the average comoving density
   distribution $n_c(r)$ as a function of the comoving distance
   $r$, the Limber equation can be used to approximately describe the expected
 cross-correlation with the galaxy distribution \footnote{The error in the Limber
 approximation, in the form expressed here, is about $10\%$ at
 $\ell=2$ and drops as $\ell^{-2}$.} (see \citep{ALS} for detailed derivation)
 \bea
  &&C_{gT} (\ell) \equiv \langle \delta^{\rm 2D}_{g, \ell m}
  T^*_{\ell m}\rangle \nonumber\\
   &&= \frac{2~ T}{\int dr ~r^2 n_c(r)}\int dr ~n_c(r) P_{\Phi^{\prime},
   g}\left(\frac{\ell+1/2}{r}\right), \label{crossform}
 \eea
  where $\delta^{\rm 2D}_{g, \ell m}$ and
  $T_{\ell m}$ are the projected survey galaxy overdensity and the CMB
  temperature in the spherical harmonic space. If we assume that the galaxies follow the matter density
  with constant bias, $b_g$, (i.e.
   $\delta_g = b_g \delta_m$), then  $P_{\Phi^{\prime},
   g}(k)$,
   the 3D cross-power spectrum of $\Phi^{\prime}$ and
   galaxy overdensity for the comoving wave-number $k$, can be related directly to $P_{\Phi^{\prime},
   m}$. We then use the $G_{00}$ Einstein equation \citep{MFB}
\beq (k^2+3{\cal H}^2)\Phi+3{\cal H}\Phi^\prime+4\pi G a^2(\rho_m
\delta_m +\rho_{DE} \delta_{DE}) =0, \label{G00}\eeq
    to relate the matter auto-power spectrum $P_{m,m}(k) = P(k)$ to $P_{\Phi^{\prime},
   g}$. Here, ${\cal H} = d\ln a/d\eta$ is the conformal Hubble constant,
   and $\rho_{DE}$ and $\delta_{DE}$ are the average density and
   overdensity of the dark energy, respectively.

   Note that, for a
   cosmological constant (a \lcdm cosmology), $\delta_{DE}=0$, and Eq. (\ref{G00})
   reduces to the Poisson equation for $k \gg {\cal H}$. In this
   case, Eq. (\ref{crossform}) reduces to
   \bea
   &&C_{gT}(\ell) =
   - \frac{3 b_g H^2_0 \Omega_m}{\int dr~r^2
   n_c(r)}\times \nonumber\\&&\int dr ~r^2~n_c(r)\cdot\frac{1+z}{(\ell+1/2)^2}\cdot
   \frac{g^{\prime}}{g}
   P\left(\frac{\ell+1/2}{r}\right),\label{cross}
   \eea
    where $g$ is the linear growth factor of the gravitational
      potential,$\Phi$, and $g^{\prime}$ is its derivative with
     respect to the conformal time, while $z$ denotes the cosmological
     redshift, which is related to the comoving distance $r$ via
     \beq
     \frac{dz}{1+z} = {\cal H} dr.
     \eeq

   For any alternative theory of dark energy, an independent equation for
the evolution of $\delta_{DE}$ should be solved simultaneously.

\section{The Error in ISW Detection}
 It is easy to see that the
expected dispersion (see \citep{ALS} for details
     \footnote{If the survey does not cover the whole sky, the nearby multipoles
     in the harmonic space will not be independent. However, if we bin a large number of multipoles together,
      we can use the increased standard deviation of Eq. (\ref{noise}) for each multipole and treat them as
      independent to estimate the error of each multipole bin.})
      in the cross-correlation signal for harmonic multipole
     $C_{gT}(\ell)$ is given by
     \beq
     \Delta C^2_{gT}(\ell) \simeq \frac{C_{gg}(\ell) C_{TT}(\ell)}{f_{\rm
     sky}(2\ell+1)}  \label{noise},
     \eeq
     where $f_{\rm sky}$ is the fraction of sky covered in the survey, and
     we assumed a small cross-correlation signal, i.e.
     $
     C^2_{gT}(\ell) \ll C_{gg}(\ell) C_{TT}(\ell),
     $
     which is the case for the ISW effect (the expected ISW effect is much smaller than the primary anisotropies,
     but see \citep{kamion}).

\begin{figure}
       \includegraphics[width=\linewidth]{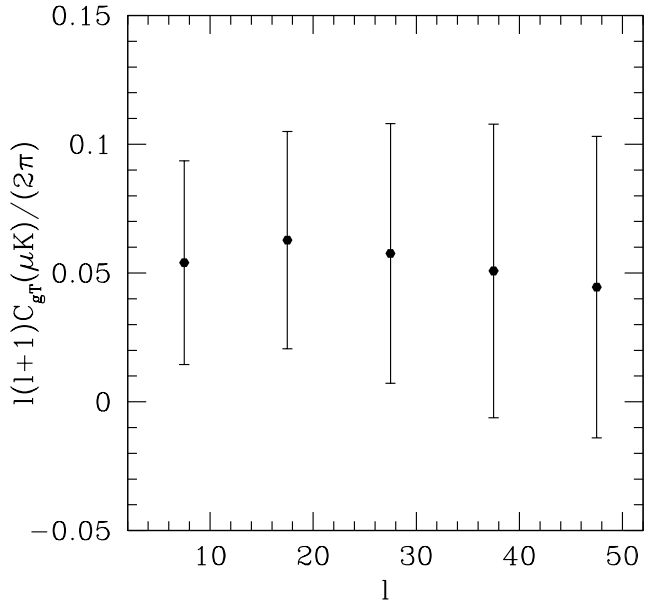}
      \caption{\label{signal} The expected cross-power spectrum of the ISW effect, for an all-sky survey with $b^2_g dN/dz =10^7$, and
      $z_{\rm max} =1$}
  \end{figure}
     $C_{TT}(\ell)$ is the observed CMB temperature auto-power, which includes both the
     intrinsic CMB fluctuations and the detector noise. As will become clear later on, since the ISW
     effect is observed at small $\ell$, the WMAP observed auto-power
     spectrum\citep{WMAPpower}, which has negligible detector noise at low
     $\ell$, should give the optimum power spectrum to use in
     Eq. (\ref{noise}). We again use the Limber approximation to obtain the projected
galaxy auto-power
     \beq
      C_{gg}(\ell) \simeq \frac{\int dr~r^2~ n^2_c(r) [b^2_g(r) \cdot
      P\left(\frac{\ell+1/2}{r}\right)+ n^{-1}_c(r)]}{\left[\int
      dr~ r^2~n_c(r)\right]^2}.\label{gal_auto}
     \eeq

        Fig. (\ref{signal}) shows an example of the
        expected cross-power signal and error for a survey with 10
        million galaxies with $b_g=1$, uniformly distributed between $0<z<1$.

\begin{figure}
       \includegraphics[width=\linewidth]{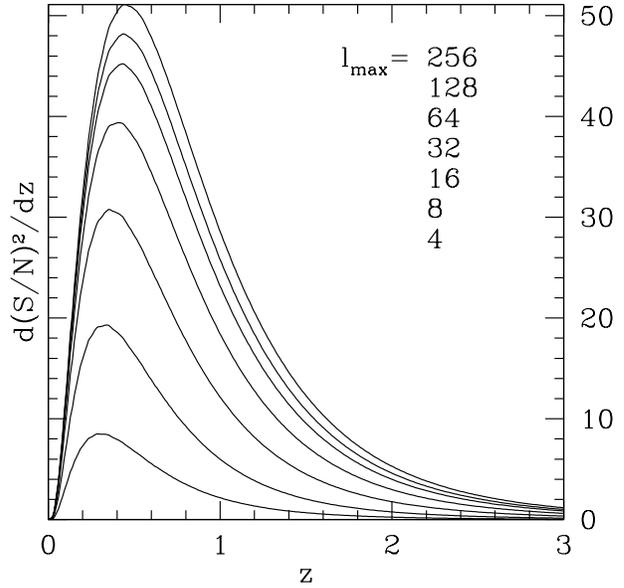}
       \includegraphics[width=\linewidth]{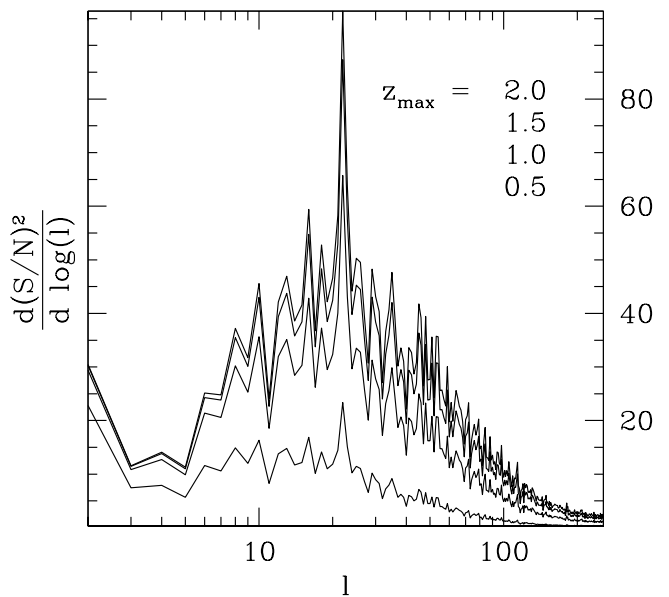}

      \caption{\label{figsnlz}Figures show the expected $(S/N)^2$ distribution for a redshift or resolution limited
      full-sky ISW
      cross-correlation signal ($\ell_{\rm max}$ refers to the scale at which the (white) detector noise is equal
      to the true signal).
      In either figures, the enclosed area for the region covered by a
      survey, multiplied by its sky coverage,
      gives the optimum $(S/N)^2$ for the cross-correlation signal. The spikiness of the
      distribution in $\ell$-space is due to the use of the actual
      observed WMAP power spectrum in Eq. (\ref{dsn21}).
      Note that for partial sky coverage,
      the low-$\ell$ multipoles that are not covered by the survey should also be excluded from the
      area under the curves.}
  \end{figure}

\section{The Perfect Galaxy Survey}\label{perfect}
     To obtain the optimum signal-to-noise ratio for an ISW detection in
     cross-correlation, we assume that we have an (at least approximate)
     redshift estimate for each galaxy in the survey. Then we can
     divide the survey into almost independent shells of thickness $\delta r$,
     where
     $$
       r_0 \ll \delta r \ll r,
     $$
     where $r_0 \lesssim 5 h^{-1} \mpc$ is the galaxy
     auto-correlation length.
         Using Eq. (\ref{crossform}), the expected ISW cross-power signal for a
         narrow redshift bin $(z,z+\delta z)$, at multipole $\ell$, is
         \beq
         C_{gT}(\ell,z) = \frac{2 b_g(r)}{r^2(z)}
         P_{\Phi^\prime,m} \left(\frac{\ell+1/2}{r}\right),\label{cgtz}
         \eeq
         while Eq.(\ref{noise}) gives the error
         in the cross-power spectrum
         \beq
         \Delta C_{gT}^2(\ell,z) =
         \frac{C_{TT}(\ell) C_{gg}(\ell,z) }{2\ell+1}\label{dcgtz},
         \eeq
         and Eq.(\ref{gal_auto}) can be used to find $C_{gg}(\ell,z)$
         \beq
         C_{gg}(\ell,z) = \frac{b^2_g(r)}{r^2 \delta r}\left[P\left(\frac{\ell+1/2}{r}\right)+(n_c b^2_g)^{-1}\right].
         \label{cgglz}
         \eeq

     Dividing Eq. (\ref{cgtz}) by Eq. (\ref{dcgtz}), we find the expected signal-to-noise ratio for
     the cross-correlation signal in multipole $\ell$, due to this shell
     \beq
     \delta (S/N)^2 = \frac{C^2_{gT}(\ell)}{\Delta
     C^2_{gT}(\ell)}
     =
     \frac{\left[  r^{-2} \delta r \cdot (2\ell+1) \right]
     \times 4 P_{\Phi^{\prime},m}^2(k)}{C_{TT}(\ell)[P(k)+(n_c
     b^2_g)^{-1}]},\label{dsn21}
     \eeq
     where $k = \frac{\ell+1/2}{r}$.
      Within the approximation of independent shells and
      multipoles, $(S/N)^2$ is cumulative, and we could simply add
      (or integrate over) the contribution due to different multipoles and shells that
      are included in the galaxy survey, and multiply it by the sky coverage, $f_{\rm sky}$,
      to obtain the optimum $(S/N)^2$ (in the absence of systematics)
      for the whole survey. Fig. (\ref{figsnlz}) shows the
      $(S/N)^2$ density distribution for hypothetical all-sky surveys
      with limited (CMB) resolution, or redshift depth, in a \lcdm cosmology, while we assumed that the
      Poisson noise is negligible ($n_c \rightarrow \infty$).

       We see that the ISW cross-correlation signal is widely distributed over a
       redshift range between $0$ and $1.5$ (which is the era in which dark energy becomes cosmologically important),
       and peaks at $z\simeq 0.4$, below which the detection is limited by the available volume.
       Almost all the signal is due to multipoles with $\ell \lesssim 100$ ($\theta \gtrsim 2^{\circ}$), which
       implies that the WMAP 1st year all-sky temperature map \citep{WMAP}, with $S/N \gg 1$ on scales larger than
       a degree ($\ell<200$), captures almost all the ISW signal in the CMB sky, while the upcoming 2nd year WMAP data,
       or higher resolution CMB observations are
       unlikely to make a significant difference. The
       angular scale of the cross-correlation signal decreases
       with the depth of the survey. This is due to the fact
       that the angular correlation length of the galaxy
       distribution is smaller for a deeper (more distant) sample.

        For our assumed \lcdm cosmological model, the total $S/N$ which could be achieved by a perfect survey is
        $\sim 7.5$. Let us compare this result with previous such estimates.
        The first estimate of $S/N$ for a perfect
        survey \citep{turok} is, in fact, completely consistent with $7.5\sigma$. Other such estimates
        \citep{ISWs} focus on specific surveys, but seem to be
        roughly consistent with the values expected from Fig.
        (\ref{figsnlz}) for those surveys. This is despite our use of Limber (Eq. \ref{crossform}) and
        independent redshift bins (Eq. \ref{dsn2}) approximations, and demonstrates that these approximations are
        adequate for our purpose. Our approach allows us to visualize where in
        redshift-angular space one should look for the ISW
        signal. However, given a sample of galaxies in existence, we should emphasize that neither we, nor previous
        works, provide an optimum procedure to estimate this signal.
        For a realistic galaxy survey, such procedure would depend on
        possible survey contaminations or systematics, detailed geometry, and redshift
        information available for the survey (See the next
        section).

\section{Sub-Optimal Galaxy Surveys}

        In this section we look into how different features of
        realistic surveys may reduce the optimum ISW
        signal-to-noise obtained in the last section.

\subsection{Poisson Limited Surveys}
        For a realistic survey, an additional source of noise is
        Poisson fluctuations in the galaxy number density. The
        Poisson noise (the second term in the brackets in Eq.
        \ref{gal_auto}) is inversely proportional to the average
        number density of observed galaxies in the survey, which
        dominates the uncertainty in cross-correlation for a small galaxy sample.
\begin{figure}
\includegraphics[width=\linewidth]{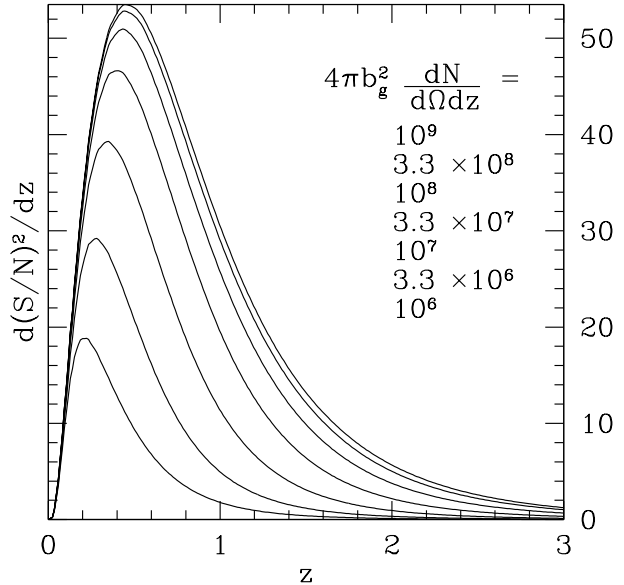}
\caption{\label{figdsnp}The distribution of $(S/N)^2$ for
different galaxy redshift distributions. For partial sky coverage,
the result should be multiplied by $f_{\rm sky}$.}
\end{figure}
\begin{figure}
\includegraphics[width=\linewidth]{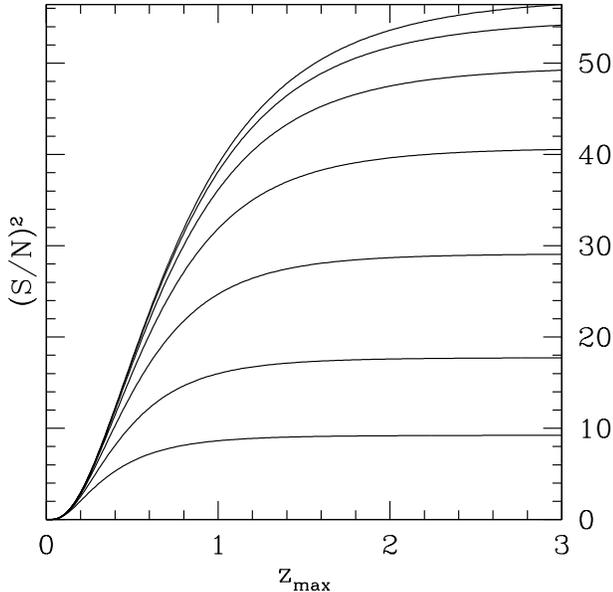}
\caption{\label{figsnp} The integrated (total) $(S/N)^2$ for a
constant $dN/dz$ up to $z_{\rm max}$, corresponding to the area
under the curves in Fig. \ref{figdsnp}.}
\end{figure}

        Figs. (\ref{figdsnp}-\ref{figsnp}) show how the total cross-correlation $(S/N)^2$ and its
        distribution depend on the number of galaxies in an all-sky
        survey. Although a fixed number of galaxies per unit redshift
        is assumed, different curves in Fig. (\ref{figdsnp}) can be combined to obtain the
        signal-to-noise for an arbitrary redshift/bias distribution.

        Fig. (\ref{figsnp}) shows that an ambitious all-sky survey with about 10 million
        galaxies (or one million clusters with $b_g\sim 3$), which uniformly cover the redshift range between 0 to
        1, can only yield a $S/N$ of $\simeq 5$. Although the Sloan
        Digital Sky Survey (SDSS; \citep{SDSS}) is unlikely to get a $S/N$
        better than $4\sigma$ due to its incomplete redshift and sky coverage,
        future galaxy surveys like LSST and Pan-STARRS could achieve a cosmic variance limited detection of ISW in
        cross-correlation \citep{survey}, should they cover almost all of
        sky, and have sufficiently small systematic
        errors (see Sec. \ref{contaminant}).

\subsection{Observational Contamination}\label{contaminant}

        Another feature of a realistic galaxy survey is the presence of
        artificial structure in the sky maps. Such structures may be
        due to observational systematics, such as day to day
        changes of weather, seeing or photometric calibrations.
        Another possibility is the effective change of magnitude limit due to Galactic
        extinction, or Galactic contamination due to stars
        misidentified as galaxies. Depending on if these
        contaminations are correlated with possible contaminations
        of the CMB maps, they can add extra random or
        systematic errors to the optimum error that may be
        achieved by a survey.

\subsubsection{Uncorrelated Contaminations: Random Errors}\label{unc}
\begin{figure}
\includegraphics[width=\linewidth]{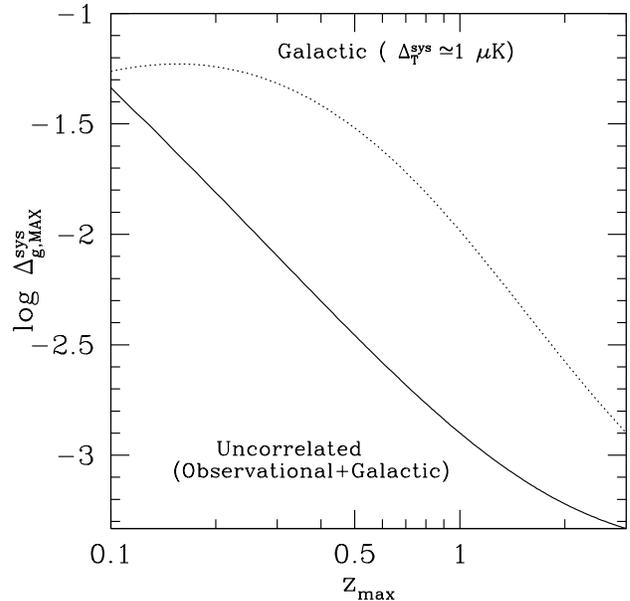}
\caption{\label{figsys} The graph shows the maximum allowed
amplitude of anisotropies, $\Delta^{sys}_g$ introduced by possible
systematics of the galaxy survey. The solid line shows the level
of systematics that reduces the $(S/N)^2$ for an ISW detection by
50\%. The dotted line shows the level of {\it correlated}
contamination that introduces a systematic error in an ISW
detection comparable to the cosmic variance. Here, we assume that
the correlated amplitude of CMB contamination is $\sim 1 \mu K$,
consistent with the expected leftover Galactic contamination in
WMAP foreground-cleaned W-band map \citep{Bennett:2003ca}.}
\end{figure}
         If the contaminations of the galaxy survey is not correlated
         with the contaminations in the CMB map, they should not affect the
         expectation value of the
         cross-correlation function of the galaxy survey with the
         CMB. We do not expect the observational systematics of the
         galaxy/CMB surveys to be
         correlated. Therefore, the only effect will be to
         increase the random error through increasing the auto-powers
         of each survey (Eq. \ref{noise}). Since the
         Galactic/systematic contributions to the power spectrum of the CMB
         are believed to be small and under control
         \citep{WMAPpower,Bennett:2003ca}, we focus on the contamination of the galaxy survey.

         We again divide our galaxy sample into independent redshift bins.
         The harmonic components of the anisotropies in each redshift bin can be divided into real and systematic parts:
         \beq
         \tilde{\delta}_g(\ell,m;z) = \delta_g(\ell,m;z)+\delta^{sys}_g(\ell,m).
         \eeq
         While the systematic contaminations, $\delta^{sys}_g(\ell,m)$, is also, in general, a function of redshift,
         for simplicity, we ignore this dependence. Of course, in practice, the level of systematics in the redshift
         range that contains the bulk of the ISW signal is the relevant quantity (see Figs. \ref{figsnlz}-\ref{figdsnp}).
         Now, the contaminated auto-power spectrum will be
         \beq
         \tilde{C}_{gg}(\ell;z_1,z_2) = C_{gg}(\ell;z_1) \delta_{z_1,z_2} + C^{sys}_{gg}(\ell),\label{cggsys}
         \eeq
         where $C_{gg}(\ell,z)$ was defined in Eq. (\ref{cgglz}), and we ignore the cross-term,
         as we do not expect any correlation between the survey systematics
         and the intrinsic fluctuations in the galaxy density. Note that, unlike the previous sections,
         different redshift bins are now correlated,
         and we will need to invert the full covariance matrix to estimate the expected signal to noise.
         Eq. (\ref{noise}) can be generalized to get the covariance matrix of the cross-power spectrum for $f_{sky}=1$
         \bea
         {\bf \tilde{C}}_{\ell; z_1,z_2} = \langle \Delta C_{gT}(\ell, z_1) \Delta C_{gT}(\ell, z_2) \rangle
         \nonumber \\= \frac{\tilde{C}_{gg}(\ell;z_1,z_2) C_{TT}(\ell)}{2\ell+1}.
         \eea
          For a low level of systematics, we can use a Taylor expansion to invert the covariance matrix
          \beq
          {\bf \tilde{C}}^{-1} \simeq  {\bf C}^{-1} - {\bf C}^{-1} {\bf C}^{sys} {\bf C}^{-1}.\label{invert}
          \eeq

          Now combining Eqs. (\ref{cgtz},\ref{dcgtz}, and \ref{cggsys}-\ref{invert}) yields an approximate
          expression for the suppression of $(S/N)^2$ due to the presence of contaminations:
          \bea
          &&\delta (S/N)^2 \simeq
           -\sum_{z_1,z_2,\ell} (2\ell+1) \frac{C^{sys}_{gg}(\ell) C_{gT}(\ell,z_1) C_{gT}(\ell,z_2)}
          {C_{TT}(\ell) C_{gg}(\ell,z_1) C_{gg}(\ell,z_2)} \nonumber \\
          && = -\sum_\ell (2\ell+1) \frac{C^{sys}_{gg}(\ell)}{C_{TT}(\ell)}\left(\sum_z \frac{C_{gT}(\ell,z)}
          {C_{gg}(\ell,z)}\right)^2.\label{dsn2}
          \eea

          Let us define the amplitude of the systematic contaminations, $\Delta^{sys}_g$ as
          \beq
          \Delta^{sys}_g \equiv  \sqrt{ \frac{\ell (\ell+1) C^{sys}_{gg}(\ell)}{2\pi}}.
          \eeq
          Again, for simplicity, we assume that $\Delta^{sys}_g$ is independent of $\ell$. Of course,
          the relevant amplitude for $\Delta^{sys}_g$
          corresponds to the $\ell$-range that contains the bulk of the ISW signal, i.e.
          $\ell \sim 20$ (see Fig. \ref{figsnlz}). The solid line in Fig. (\ref{figsys}) shows the amplitude of
          uncorrelated contamination, required to reduce the $(S/N)^2$ by 50\%.
          Thus, we see that since we need $z_{max} \gtrsim 1$ to capture most of the ISW signal (see Sec. \ref{perfect}),
          the level of systematic anisotropy in the galaxy survey should be less than
          0.1\% on scales of $\ell \sim 20$.

          As an example, let us assume the systematic anisotropy in an optical survey
          is due to uncertainties in the magnitude
          calibration. Furthermore, let us assume that we know the (approximate) redshifts, and
          in a given redshift bin, the luminosity function is given by the Schechter function \citep{Schechter}:
          \beq
          \phi(L)~ dL = \phi^* \cdot (L/L^*)^\alpha \exp(-L/L^*) (dL/L^*)
          \eeq
          where $\phi^*$, $L^*$, and $\alpha$ are characteristic galaxy number density, luminosity, and slope
          of the luminosity function, respectively. Assuming that the survey is almost complete in the given
          redshift bin, i.e. $L_{min} \ll L^*$, the systematic anisotropy is given by:
          \beq
          \Delta^{sys}_g = \frac{\phi(L) \Delta L_{min}}{\int \phi(L) dL} =
          (L_{min}/L^*)^{1+\alpha}\frac{\Delta L_{min} /L_{min}}{\Gamma(1+\alpha)}.
          \eeq
          For $\alpha \simeq -0.9$ \citep{blanton}, we can ignore $1+\alpha$ in the exponent, and use
          $\Delta L/L = 0.4 \ln(10) \Delta m$, to arrive at
          \beq
            \Delta^{sys}_g \simeq \frac{0.4 ~ \ln 10~ \Delta m}{\Gamma(1+\alpha)} \simeq 0.1 ~ \Delta m < 10^{-3},
          \eeq
          where $\Delta m$ is the uncertainty in the magnitude calibration. Thus, we see that
          $\Delta m$ needs to be $\lesssim 0.01$, for the error to be dominated by cosmic
          variance. The accuracy of this calibration is currently $\sim 0.02$ for SDSS
          \citep{Abazajian:2003jy}, and will remain a challenge for
          future surveys such as Pan-STARRS and LSST. Note that the magnitude uncertainty of individual
          galaxies, as long as they are not correlated on the scale of a few degrees on the sky,
          can be significantly larger.

          Repeating this estimate for an incomplete survey. i.e. $L_{min} > L^*$, we see that a significantly higher
          precision of magnitude measurement will be required to capture the ISW signal:
          \beq
           \Delta m \lesssim \left\{\begin{array}{ll} 0.01 & \mbox{~for $L_{min} \ll L^*$,} \\
          10^{-3} \frac{L^*}{L_{min}} & \mbox{~for $L_{min} \gg L^*$.} \end{array} \right.
          \eeq
\subsubsection{Galactic Contaminations: Systematic Errors }

          As the Galactic sources of contamination in the CMB and galaxy surveys are correlated,
          they may cause a systematic over/underestimate of the cross-correlation signal.
          In order to model this effect, we assume that this correlated signal follows the
          Schlegel, Finkbeiner, and Davis (SFD)
          Galactic extinction map \citep{Schlegel}. Since the bulk of the ISW signal comes from $\ell \sim 20$,
          we normalize the amplitude of the contaminating signals at $\ell = 20$.
          The level of Galactic contamination in the foreground cleaned W-band of the WMAP
          satellite \citep{Bennett:2003ca} is expected to be $\sim 1 ~\mu K$ outside the Galactic plane,
          and thus we take
          \bea
          \Delta^{sys}_T (\ell) \equiv \sqrt{ \frac{\ell (\ell+1) C^{sys}_{gg}(\ell)}{2\pi}} = 1 ~\mu K,\nonumber \\
           {\rm for} ~ \ell = 20,
          \eea
           while we assume $C^{sys}_{gg} \propto C^{sys}_{TT} \propto C_{\rm SFD}$,
           for the $\ell$-dependence of the correlated contamination. Here, $C_{\rm SFD}$ is the
           angular auto-power spectrum of the SFD extinction map.

           The dotted line in Fig. (\ref{figsys}) shows the amplitude of the correlated
           systematic anisotropy, normalized at $\ell = 20$, which yields a systematic error in the ISW signal
           equal to the cosmic variance error. Since this systematic error is proportional to
           $\Delta^{sys}_T \times \Delta^{sys}_g$, the upper bound, $\Delta^{sys}_{g,\rm MAX}$,
           shown in this figure decreases for larger Galactic contaminations in the CMB,
           inversely proportional to $\Delta^{sys}_T$. Thus, we see that since the solid curve is below the dotted
           curve, as long as $\Delta^{sys}_T < ~{\rm few}~ \mu K$, the level of uncorrelated contamination puts a more
           stringent constraint on the systematics of the galaxy survey. As we discussed in Sec. \ref{unc}, this
           requires the systematic anisotropy of the survey to be less than 0.1 \%.

\begin{figure}
        \includegraphics[width=0.45\linewidth]{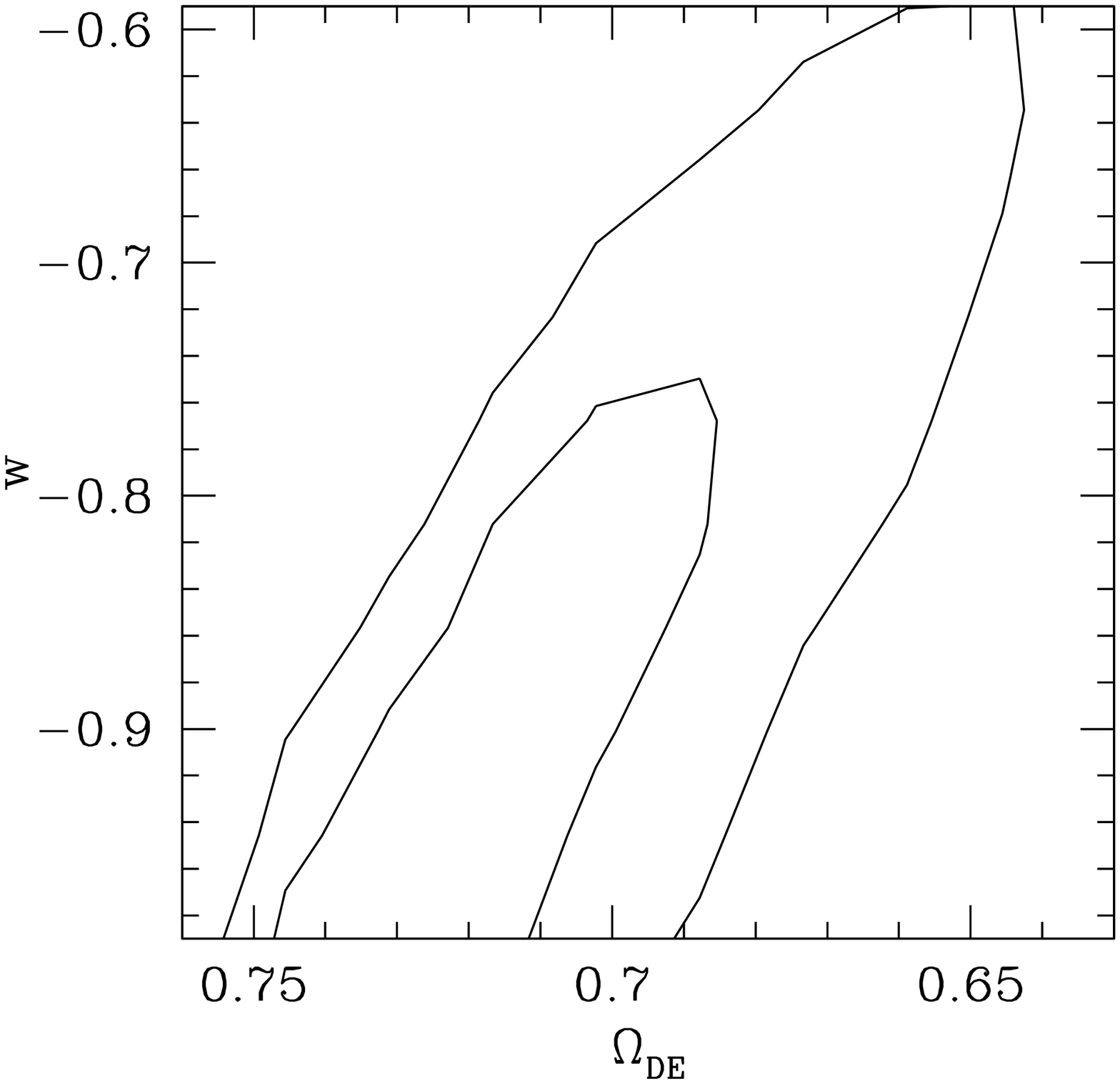}
      \includegraphics[width=0.45\linewidth]{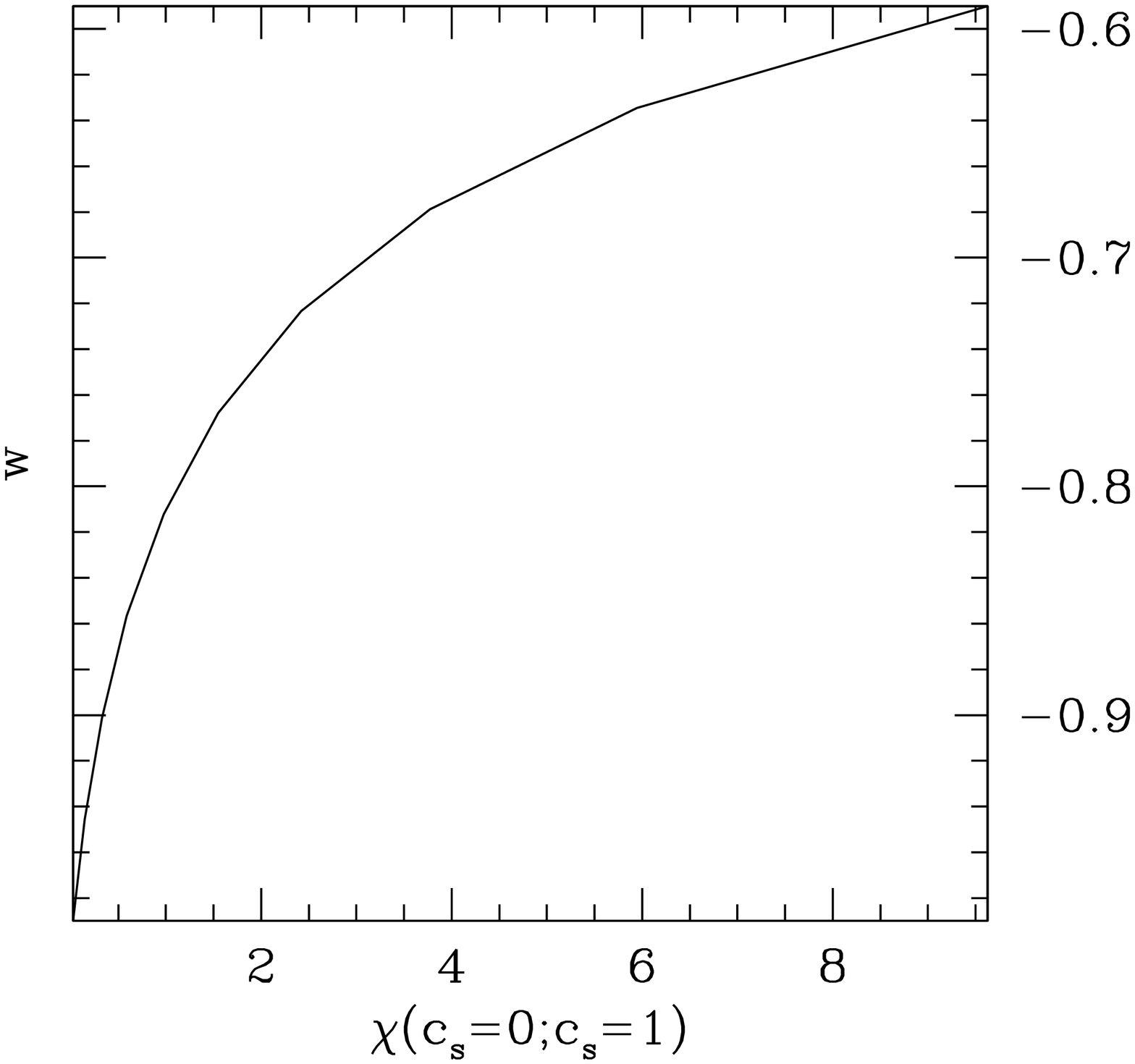}
      \caption{\label{figchi}(Left) The $w-\Omega_m$ constraints (68\% and 95\% contours) based on
      an optimum ISW detection. (Right) The significance of ruling out $c_s=0$, for a quintessence
model with given $w$, and $c_s=1$,
      based on an optimum ISW detection. In both graphs, other cosmological parameters are kept constant.}
      \end{figure}
\subsection{Redshift errors}

            Since the ISW kernel has such a broad distribution
            in redshift space, redshift errors are not expected to be a big limiting factor for ISW
            cross-correlation measurements.

            We can simulate {\it random} errors in the estimated
            redshifts, via re-writing Eq. (\ref{dsn2}) for finite redshift bins of thickness $2 \Delta z$.
            It turns out that even using bins as thick as $\Delta z \sim 0.5$ will not decrease
            the signal-to-noise by more than 5\%.

            {\it Systematic} errors in redshifts, or in redshift distribution, may also systematically
             bias the estimates of ISW amplitude. Taking the unbiased ISW amplitude to be one,
             the ISW amplitude, biased by the systematic redshift error $\Delta z$ is given by
             \bea
              1+\frac{\delta {\rm ISW}}{\rm ISW} = \sum_{\ell,z} \frac{C_{gT}(\ell,z+\Delta z)/C_{gT}(\ell,z)}
              {\Delta C^2_{gT}(\ell,z)/C^2_{gT}(\ell,z)} \nonumber\\  {\big /} \sum_{\ell,z} \frac{1}
              {\Delta C^2_{gT}(\ell,z)/C^2_{gT}(\ell,z)},
             \eea
              which, for small $\Delta z$, yields
             \bea
                \frac{|\delta {\rm ISW}|}{\rm ISW} &\simeq & \nonumber \\ \frac{1}{2}\left|\sum_{\ell,z} \frac{\Delta z \cdot
                \partial C^2_{gT}(\ell,z)/\partial z}{\Delta
                C^2_{gT}(\ell,z)}\right|
                &{\huge /}& \sum_{\ell,z} \frac{C^2_{gT}(\ell,z)}{\Delta C^2_{gT}(\ell,z)} \nonumber \\
                &\lesssim 3 ~\Delta z,&
             \eea
              where we used Eqs. (\ref{cgtz}-\ref{cgglz}) to estimate the numerical value of the above expression.
              Since the optimum signal-to-noise for an ISW detection is $\sim 7$, the redshift systematic errors may not
              dominate the errors if
              \beq
              \Delta z_{sys} \lesssim \frac{1}{3 \times 7} \simeq 0.05.
              \eeq

\section{What does ISW tell us about Cosmology?}
         Let us study the optimum constraints that
         an ISW cross-correlation detection can give us about
         cosmology. For a nominal concordance cosmology ${\cal
             C}_0$, the expected significance level for ruling out the
             cosmology ${\cal C}$, $\chi$, is given by
         \beq
             \chi^2 = \sum_{\ell,z} \frac{[C_{gT}(\ell,z;{\cal C})-C_{gT}(\ell,z;{\cal
             C}_0)]^2}{\Delta C_{gT}^2(\ell,z;{\cal
             C}_0)},\label{chi2}
          \eeq
            where $C_{gT}$ and $\Delta C_{gT}$ are defined in Eqs. (\ref{cgtz}-\ref{cgglz}).
            Note that the bias factor, $b_g$, is cancelled from the
            numerator and the denominator, and the optimum
            significance level only depends on the fluctuations in
            matter density and gravitational potential.

            Fig. (\ref{figchi}) shows the
            optimum constraints that an ISW detection may
            yield on some of the properties of dark energy, i.e. its density, equation of state,
            and speed of sound \citep{erikson}. While such constraints are already
            comparable to the current bounds on these
            parameters \citep{SpergelEtAl2003,WMAPSDSS,BeanDore},
            future observations of CMB and large scale structure
            \citep{future} will significantly improve these
            bounds. However, ISW constraints could still be
            used to test possible systematics that may affect
            the accuracy of future measurements.

            It has been claimed \citep{Garriga} that the cross-correlation
            observations could be used to break the degeneracies in
            the so-called doomsday cosmological scenarios, which
            involve a varying equation of state. However, given the
            cosmic variance limited measurements of the ISW effect
            (Eq. \ref{chi2}), it does not seem that such
            constraints can significantly improve the expected
            combined constraints from the upcoming Planck and SNAP satellite
            experiments \citep{jan1,jan2}

\begin{figure}
       \includegraphics[width=\linewidth]{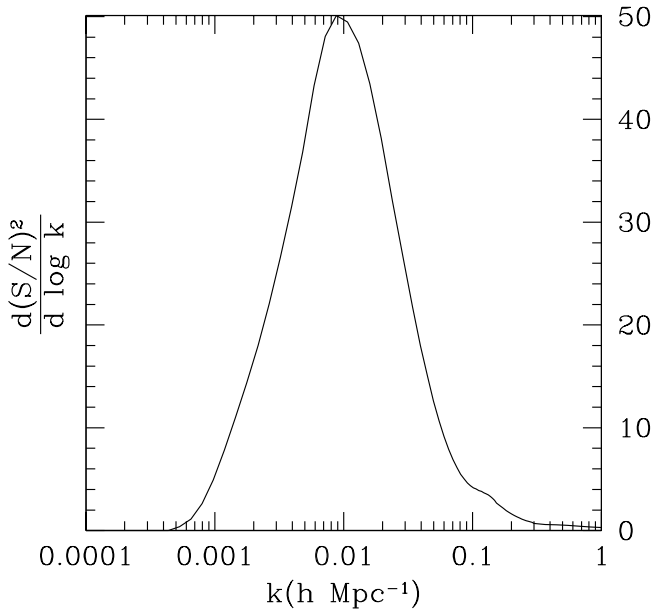}
      \caption{\label{kspace} The distribution of $(S/N)^2$ of an ISW detection in k-space.}
  \end{figure}

            A more intriguing application of an ISW detection is
            the possibility of testing our theories of matter/gravity on
            large scales \citep{lue}. While the consistency of the power spectrum of 2dF and SDSS
             galaxy surveys \citep{power} with the WMAP power spectrum confirms
            the consistency of the \lcdm concordance cosmology at
            scales of $k\gtrsim 0.1 ~h ~{\rm Mpc^{-1}}$, Fig.(\ref{kspace}) shows that an ISW detection
            can do the same at scales of $k\sim 0.003-0.03 ~h~ {\rm
            Mpc^{-1}}$. Therefore, current \citep{crosses,ALS}
            and future observations of an ISW effect may confirm the
            consistency of our cosmological theories at the largest physical scales that
            they have ever been tested. While the
            cross-correlation statistics are almost free of systematic
            bias, the auto-correlation is often dominated
            by survey systematics on such scales.

\section{Conclusions}

           In this paper, we study different aspects of the
           correlation between a galaxy survey and CMB sky, due to the
           ISW effect in a flat universe. The main source of noise is contamination
           by the primary CMB anisotropies. We see that, given this noise, most of
           the signal comes from $\ell \sim 20$, and $z\sim 0.4$
           with negligible contribution from $\ell >100$ and
           $z>1.5$.

           While the maximum significance level for an ISW detection is
           $\sim 7.5\sigma$ for a concordance \lcdm cosmology, an all-sky survey
           with about 10 million galaxies within $0<z<1$ should yield an almost perfect ISW
           detection, at the $\sim 5\sigma$ level.

           We show that,
           in order to achieve a cosmic-variance limited ISW
           detection, the systematic anisotropies of the galaxy
           survey must be below $0.1 \%$, on the scale of $\sim
           10$ degrees on the sky. Within this level of systematics, and given the current level of
           Galactic contamination in WMAP CMB maps ($\sim \mu K$),
           there should be a negligible over/underestimate of the
           ISW signal due to the systematic correlations of CMB
           and galaxy maps. We also argue that, while the random
           uncertainties in redshifts cannot significantly reduce
           the significance of an ISW detection, the systematic
           errors in redshift estimates need be less than $0.05$,
           in order to achieve a cosmic-variance limited
           measurement.

            It turns out that, due to the large noise induced
            by the primary anisotropies, the optimum constraints
            on the properties of dark energy from an ISW
            detection, are already comparable to the current
            accuracies, and will be outdone by future
            observations. However, the simplicity of the linear physics
            involved in the ISW effect, and the fact that the
            cross-correlation statistics are not biased by the
            systematics of CMB or galaxy surveys, makes ISW
            detection a useful indicator of possible systematics in
            more accurate methods.

            Finally, we point out that the detection of the ISW effect provides a
            unique test of our concordance cosmological model on the largest
            physical scales that it has ever been tested.

            \acknowledgements The author wishes to thank  Steve Boughn, Yeong-Shang Loh, David
            Spergel, and Michael Strauss for illuminating discussions and useful
            comments. This work was originated as response to a question asked by
            Yeong-Shang Loh, and the author is grateful to him for asking that question.

\end{document}